\documentclass[conference]{IEEEtran}
\IEEEoverridecommandlockouts
\usepackage{cite}
\usepackage{amsmath,amssymb,amsfonts}
\usepackage[ruled,vlined]{algorithm2e}
\SetKwProg{Init}{init}{}{}
\SetKwInOut{Input}{Input}
\SetKwInOut{Output}{Output}
\SetKwInOut{Parameter}{Parameter}
\usepackage{subcaption}
\usepackage{graphicx}
\usepackage{textcomp}
\usepackage{xcolor}
\def\BibTeX{{\rm B\kern-.05em{\sc i\kern-.025em b}\kern-.08em
    T\kern-.1667em\lower.7ex\hbox{E}\kern-.125emX}}

\newcommand{\bsym}[1]{\boldsymbol{#1}}

\begin{document}

\title{Online Segmented Recursive Least-Squares for Multipath Doppler Tracking}

\author{
    \IEEEauthorblockN{Jae Won Choi\IEEEauthorrefmark{1}, Girish Chowdhary\IEEEauthorrefmark{1}, Andrew C. Singer\IEEEauthorrefmark{1}, Hari Vishnu \IEEEauthorrefmark{2}, Amir Weiss\IEEEauthorrefmark{3}, Gregory W. Wornell\IEEEauthorrefmark{3}, and\\
    Grant Deane \IEEEauthorrefmark{4}}
    \IEEEauthorblockA{\IEEEauthorrefmark{1} \textit{Coordinated Science Laboratory, University of Illinois Urbana-Champaign}, Urbana, IL USA}
    \IEEEauthorblockA{\IEEEauthorrefmark{2} \textit{Acoustic Research Laboratory, National University of Singapore}, Singapore }
    \IEEEauthorblockA{\IEEEauthorrefmark{3} \textit{Electrical Engineering and Computer Science, Massachusetts Institute of Technology}, Cambridge, MA USA }
    \IEEEauthorblockA{\IEEEauthorrefmark{4} \textit{Scripps Institution of Oceanography, University of California at San Diego}, La Jolla, CA USA }
    \IEEEauthorblockA{\{choi223, girishc, acsinger\}@illinois.edu, harivishnu@gmail.com, \{amirwei,gww\}@mit.edu, gdeane@ucsd.edu}
}

\maketitle

\begin{abstract}
Underwater communication signals typically suffer from distortion due to motion-induced Doppler. Especially in shallow water environments, recovering the signal is challenging due to the time-varying Doppler effects distorting each path differently. However, conventional Doppler estimation algorithms typically model uniform Doppler across all paths and often fail to provide robust Doppler tracking in multipath environments. In this paper, we propose a dynamic programming-inspired method, called online segmented recursive least-squares (OSRLS) to sequentially estimate the time-varying non-uniform Doppler across different multipath arrivals. By approximating the non-linear time distortion as a piece-wise-linear Markov model, we formulate the problem in a dynamic programming framework known as segmented least-squares (SLS). In order to circumvent an ill-conditioned formulation, perturbations are added to the Doppler model during the linearization process. The successful operation of the algorithm is demonstrated in a simulation on a synthetic channel with time-varying non-uniform Doppler.
\end{abstract}

% \begin{IEEEkeywords}
% component, formatting, style, styling, insert
% \end{IEEEkeywords}

\section{Introduction}
\begin{figure}
    \centering
    \includegraphics[width = \linewidth]{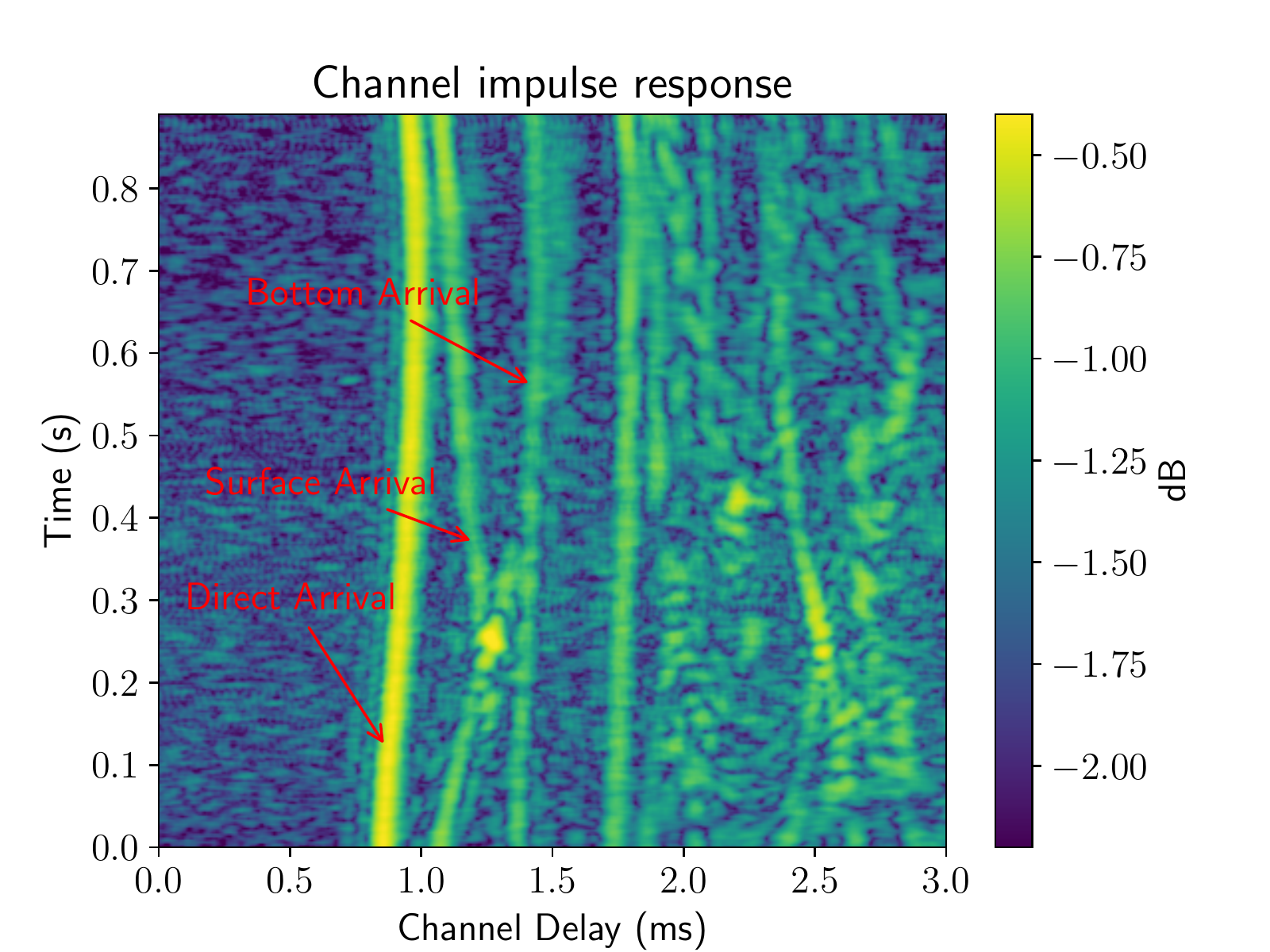}
    \caption{Channel impulse response from a single-hydrophone recording made in a wave-tank at the SOARS facility at the Scripps Institution of Oceanography. Surface waves with approximately 0.6~Hz, 0.33~m peak-to-peak amplitude and 3~m wavelength were generated during the recording. The transmitter and receiver were also moving due to the surface waves. Peaks for direct, single surface bounce, and single bottom bounce arrivals are denoted in the figure.}
    \label{fig:SOARS}
\end{figure}

Due to the motion of transmit and receive platforms as well as that of scatterers in  underwater acoustics, Doppler estimation can be a critical part of underwater signal processing applications, such as communication and localization. Ambiguity-function based methods have been used for Doppler compensation under mild conditions without substantial multipath \cite{Janus,SBA}, where gradient-based methods have also proven useful\cite{adams_dop}. For rapidly time-varying Doppler, the performance of such methods degrades rapidly. Additionally, such methods often fail to work in an environments with non-uniform Doppler across different multipath arrivals, such as shallow water in high sea states. As can be seen from Fig. \ref{fig:SOARS} based on underwater acoustic measurements taken in the presence of surface waves generated at the Scripps Ocean-Atmosphere Research Simulator (SOARS) research facility,  sensor motion, surface waves, and multiple reflections result in relative path-length variation along different multipath components in realistic scenarios, which  lead to non-uniform Doppler that would pose significant challenges for single hydrophone adaptive acoustic communication receivers.

Although there has been success in integrating Doppler compensation in acoustic communication applications in general, e.g. \cite{stoj_dop,adams_dop,willet_dop,riedl_dop}, and when a receive-array can spatially separate distinct multi-path components that contain different Doppler components such that each can be then separately tracked and compensated, \emph{sequential} tracking of individual Doppler components along different multi-path components that are not spatially separated remains an unsolved problem, such as  for single-hydrophone receivers.

One approach to resolving time-varying non-uniform Doppler is a dynamic programming approach using dynamic time-warping in \cite{DTW}. However, this is a batch algorithm computed in hindsight after all observations are available, which can lead to unacceptably long delays in communications applications. Additionally, the accuracy of such discrete dynamic time-warping methods is limited by the inherent quantization of the time-scales considered.

In this paper, we propose a sequential Doppler estimation method, inspired by a dynamic programming approach for piece-wise linear function approximation called Segmented least-squares (SLS) and an online segmented recursive least-squares variant (OSRLS) proposed in \cite{OSRLS}. The main contribution of this paper is formulating the Doppler estimation problem as an SLS problem, by approximating the non-linear time distortion via a piece-wise linear model, and linearizing the received signal with respect to the linear time-scale coefficient. 
%Further, to increase stability, a perturbation step is added during the linear approximation of the Doppler estimation problem. 
The rest of the paper is organized as follows. In Section \ref{sec:OSRLS}, we briefly review OSRLS. Section \ref{sec:form} is concerned with the linearization of the received signal model and the formulation of Doppler estimation via the SLS framework. In Section \ref{sec:sim} we demonstrate the operation of the algorithm on synthetic data consisting of a communication signal transmitted over a simulated channel with non-uniform, time-varying Doppler inspired by our observations in highly dynamic environments, such as that shown Fig. 1. Section \ref{sec:con} offers some conclusions.

\section{Online Segmented Recursive Least-Squares}\label{sec:OSRLS}
The OSRLS method is inspired by the SLS method developed by Bellman \cite{bellman} for approximating a curve with piece-wise linear segments. The goal of SLS is to minimize the objective function
\begin{align}
    E(P) = |P|C + \sum_{p_i \in P} e_{p_i} \label{eq:SLS}
\end{align}
by choosing segment set $P \triangleq \bigcup p_i$ where $\{p_i\}$ refer to non-overlapping time segments, in which each segment has a squared linear approximation error $e_{p_i}$. The constant $C$ is a penalty term for adding a segment and $|P|$ denotes the cardinality of the set $P$. 
%For each segment, we have the least-squares error (LSE) term
%\begin{align}
%    e_{p_i} = \min_x ||y_{p_i}-A_{p_i}x||^2,
%\end{align}
%for observations $y_{p_i}$ 
The Bellman equation that recursively solves for the optimal segmentation into piece-wise linear regions can be formulated with slight abuse in notation as
\begin{align}
    E(b_i) = \min_{1\leq a_i < b_i} e_{(a_i,b_i)} + C + E(a_i), \label{eq:bellman}
\end{align}
where each $(a_i,b_i)$ pair refers to the segment $p_i$ expressed in terms of its starting point and end point, respectively. Although \eqref{eq:bellman} is the optimal solution to \eqref{eq:SLS}, the solution can only be obtained in hindsight when we have all observations
%to construct $y_{p_i}$ and $A_{p_i}$ 
for all segment pairs. Also, it is computationally expensive since computation over all $(a_i,b_i)$ pairs is required.

The OSRLS leverages the approach behind the SLS partitioning but in a \emph{strongly sequential} manner. A summarized description of OSRLS is provided below. More details can be found in \cite{OSRLS}.

At each iteration $n$, the LSE is computed for all $(m, n)$ pairs for $m = 1, 2,\dots n$. The LSE and the solution for each $(m,n)$ pair can be computed recursively  using the Woodbury identity and a recursive least-squares update \cite{OSRLS}. At each $n$, the solution to \eqref{eq:bellman}, $E(n)$, is computed and compared with $E(n-1)$. If there is an increase in the error by more than a pre-chosen threshold, then the sequential algorithm (OSRLS) assumes that a new segment has started and the RLS Riccati equation is reset. In \cite{OSRLS}, sequential performance is shown to compete well with batch performance for piece-wise stationary sources.
The algorithm has a computational cost of $O(N^2)$ for a length-$N$ observation and a low-complexity variation of $O(N)$ is also shown in \cite{OSRLS}.

\section{Problem Formulation}\label{sec:form}
\subsection{Linearization of Received Signal}
We now formulate the time-varying non-uniform multipath Doppler estimation problem in the SLS framework. In a multipath and non-uniform Doppler setting, a propagated acoustic signal $s(t)$ sampled at the receiver can be expressed as
\begin{align}
    r[n] = \sum_{\ell=1}^{L} h_\ell s(\alpha_{\ell}(n)) + \omega[n],
\end{align}
where $h_\ell$ is a channel gain, $\omega[n]$ is a sampled white Gaussian process, and $\alpha_\ell(n)$ is a time distortion function of the receive time of the $n^{\text{th}}$ sample to the corresponding transmission time for each path $\ell$ amongst $L$ paths in total.

If we assume that the first derivative of $\alpha_\ell$ changes sufficiently slowly relative to the sampling rate, we can model $\alpha_\ell$ as a piece-wise linear process with respect to $n$. For each segment $p_i$ and sampling rate $\frac{1}{T}$, the time distortion function can be approximated linearly as 
\begin{align}
    \alpha_{\ell,p_i}(n) \approx d_{\ell,i}(n-a_i)T+\tau_{\ell,i}, \; n\in p_i \label{eq:part_rx}.
\end{align}
The process is characterized by the time scaling Doppler $d_{\ell,i}$ and the segment delay $\tau_{\ell,i}$ for each segment $i$ and path $\ell$. Given the initial arrival time $\tau_{\ell,0}$, the segment delays are recursively computed as
\begin{align}
\tau_{\ell,i} = \tau_{\ell,{i-1}} + d_{\ell,{i-1}}(b_{i-1}-a_{i-1})T. \label{eq:delay_update}
\end{align}
Thus, the function $\alpha_\ell$ can be parameterized by a set of Doppler factors $\{d_{\ell,i}\}$ and segments $\{p_i\}$.
Assuming the channel gains and the initial delay for each path are known, an estimate of the received signal can be computed with Doppler $d_{\ell,i}$ and the segment $p_i$ as
\begin{align}
    \hat{r}_{p_i}(n,\mathbf{d}_{i}) = \sum_{\ell=1}^L h_\ell s(d_{\ell,i}(n-a_i)T+\tau_{\ell,i}), \; n \in p_i,\label{eq:part_rx}
\end{align}
where $\mathbf{d}_{i} = [d_{1,i} \; d_{2,i} \; \dots \; d_{L,i} ]^{\rm T }$. The Doppler estimation objective now becomes an estimation of $\mathbf{d}_{i}$ and the corresponding partition $p_i$. 

Given the partition $p_i$, the respective LSE is written as
\begin{align}
    e_{p_i}= \min_{\mathbf{d_{i}}} \sum_{n=a_i}^{b_i} \left(r[n] - \hat{r}(n,\mathbf{d}_{i}) \right)^2.\label{eq:part_LSE}
\end{align}

For a small change in Doppler per segment, \eqref{eq:part_rx} can be approximated using a first-order Taylor  expansion as
\begin{align}
    \hat{r}(n,\mathbf{d}_{i} ) \approx \hat{r}(n,\mathbf{d}_{i-1} ) + \nabla_{\mathbf{d}_{i-1}}^{\rm T} \hat{r}(n,\mathbf{d}_{i-1} )\Delta\mathbf{d}_i, \label{eq:linearization}
\end{align}
for $\Delta\mathbf{d}_i = \mathbf{d}_{i}-\mathbf{d}_{i-1}$.
Given the Doppler factor of the previous partition, we can rewrite \eqref{eq:part_LSE} as 
\begin{align}
    e_{p_i}= \min_{\Delta\mathbf{d}} || \mathbf{r}_{p_i} - \mathbf{\hat{r}}_{p_i}(\mathbf{d}_{i-1}) - \mathbf{R}_{i-1}\Delta\mathbf{d}_i||^2, \label{eq:lin_LSE}
\end{align} 
where $\mathbf{r}_{p_i}$ and $\mathbf{\hat{r}}_{p_i}$ are vectors with elements $r[n]$ and $\hat{r}(n,\mathbf{d}_{i-1})$ for all $n \in p_i$, respectively, and $\mathbf{R}_{i-1}$ is the Jacobian matrix whose rows are $\{\nabla_{\mathbf{d}_{i-1}}^{\rm T} \hat{r}(n,\mathbf{d}_{i-1} \}$.
\subsection{Perturbation for Stability}
For a special case where all elements of $\mathbf{d}_{i-1}$ have the same value, $\mathbf{R}_{i-1}$ becomes a rank-1 matrix, making \eqref{eq:lin_LSE} an ill-conditioned least-squares problem. To circumvent this possible technical issue, and to ensure numerical stability, we add a small perturbation to ensure that $\mathbf{R}_{i-1}$ has full column rank. For each $n^{\text{th}}$ received sample, the first order Taylor expansions are computed with respect to $\mathbf{d}_{i-1}+\boldsymbol{\epsilon}_\ell$ for $\ell = 1, 2, \ldots, L$, where $\boldsymbol{\epsilon}_\ell$ is a vector with some small value $\epsilon$ as the $\ell^{\text{th}}$ element and zeros in all others. In this case, the perturbed linearization of \eqref{eq:linearization} is
\begin{equation} \label{eq:perturb_lin}
\begin{split}
\hat{r}(n,\mathbf{d}_{i} ) \approx  \hat{r}&(n,\mathbf{d}_{i-1}+\bsym{\epsilon}_\ell ) \\&+  \nabla_{\mathbf{d}_{i-1}+\bsym{\epsilon}_\ell}^{\rm T} \hat{r}(n,\mathbf{d}_{i-1}+\bsym{\epsilon}_\ell )(\Delta\mathbf{d}_i-\bsym{\epsilon}_\ell ),
\end{split}
\end{equation}
For each new observation, the RLS update runs $L+1$ times with the original linearization and each perturbed linearization model.
% \subsection{Penalty Term}
% The penalty term in the SLS formulation is used to capture the \textit{a priori} information about the physical limitation of the changes in Doppler. The penalty term is chosen as a function of the $\mathbf{d}_{i-1}$ as
% \begin{align}
%     C = \sigma ||\Delta\mathbf{d}_i||^2 + c,
% \end{align}
% where $\sigma$ and $c$ are design parameters of choice. The parameter $c$ is a fixed penalty term for changing the segment, and $\sigma$ is a scaling term that penalizes large change in Doppler. 

\subsection{Proposed Algorithm}
The summary of the perturbation-driven OSRLS algorithm for Doppler estimation is given in Algorithm \ref{alg:OSRLS}. The reduced (linear) complexity variation is used since in practical applications for underwater acoustic Doppler estimation, such as underwater acoustic communication, the observation length is often large. In the linear variation, $N_s$ segments with the smallest LSE and $N_r$ most recent segments are maintained. 

\begin{algorithm}
\SetAlgoLined
\Parameter{}
\quad $C \in \mathbb{Z}^{+}$ - Segment penalty term;\\
\quad $M \in \mathbb{Z}^{+}$ - Segment detection threshold;\\
\quad $N_s \in \mathbb{Z}^{+}$ - Memory for smallest LSE values;\\
\quad $N_r \in \mathbb{Z}^{+}$ - Memory for recent LSE values; \\
\quad $[\bsym{\epsilon}_1,\bsym{\epsilon}_2,\dots \bsym{\epsilon}_L]^{\rm T} \in \mathbb{R}^{L}$ - Perturbation point;\\

 \Input{}
\quad $r[n]$ - Sampled received signal;\\
\quad $s(t)$ - Transmission signal function;\\
 
 \Output{}
\quad $\mathbf{d}_1,\mathbf{d}_2,...$ - Estimated multipath Doppler;\\
\quad $p_1,p_2,...$ - Estimated segments;\\

 \Init{}{
 $i=1$;\\
 $a_1 = 0$;\tcp {Initial segment start point}
 $m_f = 0$;\tcp {\# of memory filled}
 $\mathbf{d}_0 = [1\; 1 \ldots \; 1]^{\rm T}$;\tcp {Initial Doppler}
 }
\While{$n= 1,2,...$}
{
     \tcp{Memory is full, discard one}
    \If{$m_f = N_s+N_r$} 
    {
        Discard the segment with largest LSE among $N_s$;\\
        $m_f = m_f-1$;
    }
    \tcp{Update segments in the memory}
    \While{$m=1,2,\dots m_f$}{
        \While{$\ell=1,...L$}{
            
            Linearize \eqref{eq:part_rx} with $\mathbf{d}_{i}+\boldsymbol{\epsilon}_\ell$;\\
            Solve \eqref{eq:lin_LSE} for the $(\hat{a}_m,n)$ segment using RLS to obtain $e_{\hat{a}_m,n}$ and $\Delta \mathbf{d}_{(\hat{a}_m,n)}$;
        }
    }
    \tcp{Add a new segment to the memory}
    Initialize a RLS algorithm at the segment starting at $\hat{a}_{m_f} = n-1$; \\
    $m_f=m_f+1$;\\
    
    \tcp{Segment Detection}
    Compute $E(n)$ and $a^*_n$ such that $E(n) = \min_{a^*_n \in \{ \hat{a}_1 \dots \hat{a}_{m_f}\} } (e_{a^*_n,n}+C+E(a^*_n))$;
    
    \eIf{$a^*_n -a^*_{n-1} \geq M$ }{
        $\mathbf{d}_i = \mathbf{d}_{i-1}+\Delta\mathbf{d}_i$;\\
        $p_i = (a_i,b_i)$;\\
        $i = i + 1;$\\
        $a_i = a^*_n$;\\
        $b_{i-1} = a^*_n-1 $;\\
        
        $\Delta \mathbf{d}_i = \Delta \mathbf{d}_{(a^*_n,n)}$;\\
        Reset RLS-LSE Riccati equation with RLS values from $(a^*_n,n)$ segment;\\
    }
    {
        Update $\Delta \mathbf{d}_i$ using RLS to solve for \eqref{eq:lin_LSE};
    }

 }
 \Return{$\{\mathbf{d}_i,p_i \}$};
 \caption{Online Segmented Recursive Least-Squares for Multipath Doppler Tracking}\label{alg:OSRLS}
\end{algorithm}

\section{Simulation Results}\label{sec:sim}
In this section, the OSRLS performance is evaluated for a time-varying and non-uniform Doppler using a 3-ray model \cite{3ray} (consisting of direct, surface-reflected and bottom paths) in a simulation, and is compared with a peak-tracking algorithm on multipath delays. The simulation is generated based on the settings in the SOARS experiment recordings used to obtain the channel estimates in Fig. 1. 

In our setting, a 20 kHz symbol rate quadrature phase-shift keying (QPSK) signal with Gaussian-windowed pulse shaping is used to ensure that the transmitted signal is differentiable. The signal is modulated by a carrier frequency of 30 kHz and sampled at 200 kHz. The bottom is assumed flat with a depth of 1.8 m. A transmitter and receiver are located at depths of 0.46~m, and 1.45~m away from each other. The transmitter is stationary and the receiver oscillates in the horizontal direction with a 0.6 Hz sinusoidal motion with 0.25~m peak-to-peak amplitude. The surface motion is modeled as a sinusoidal fluctuation of 0.6 Hz and 0.33 m peak-to-peak amplitude. The channel gains for direct, surface, and bottom arrivals are set to 1, -0.8, and 0.5, respectively.

For the OSRLS algorithm parameters, segment penalty $C$, segment detection threshold $M$, and the Doppler factor perturbation value $\epsilon$ are chosen here as $0.01$, $50$, and $10^{-6}$, respectively. The 20 most recent segments and 10 smallest LSE value segments (excluding the 20 most recent) are selected for the recursion at each iteration. For the peak-tracking algorithm, we obtain a cross-correlation output between the received signal and 3 ms segments of the transmitted signal for each sample iteration. At each iteration, peaks of matched filter outputs are tracked by choosing the local maxima closest to the previous peaks. Since these peaks are the maxima of a sampled signal, time-delay for each path is estimated by fitting a second-order polynomial through a maximum and two neighboring points \cite{subsample}. For both algorithms, we assume that the initial time delays for all paths are known.

\begin{figure}
    \centering
    \includegraphics[width = \linewidth]{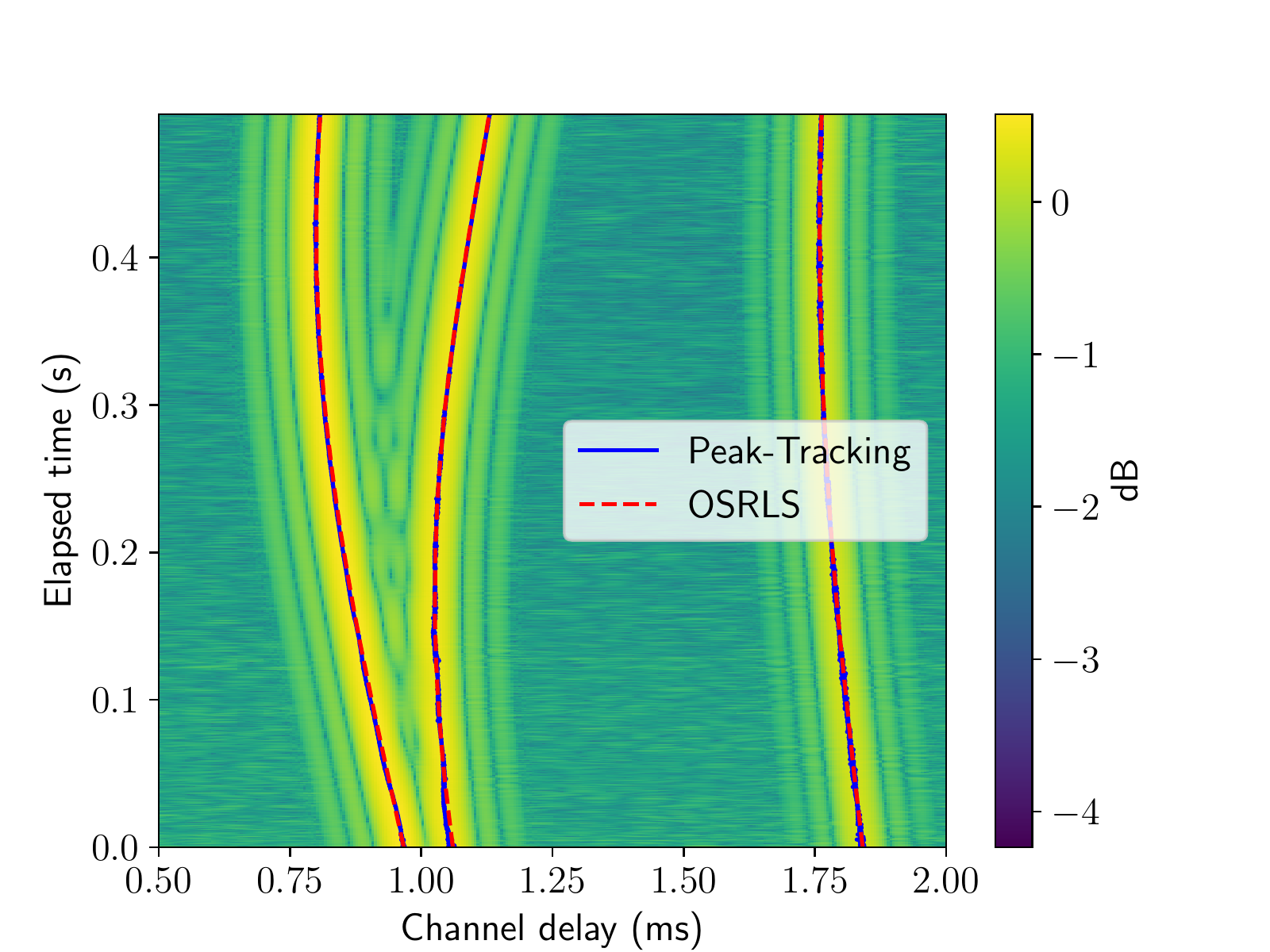}
    \caption{Time-varying Channel delay computed with Doppler factor estimated from OSRLS (red line) and peak-tracking method  (blue line) on top of channel impulse response estimate over time shown as a 2-dimensional surface plot where the vertical axis is time (in s) and horizontal axis is delay (in ms).}
    \label{fig:channel}
\end{figure}

Fig. \ref{fig:channel} illustrates the evolution of the simulated channel delay over time. The direct path and the bottom bounce are the first and third arrivals, respectively, on which we can observe similar Doppler. The second arrival is the surface reflection, which has a distinct time delay evolution different from the other two arrivals due to the surface motion.
\begin{figure}
    \centering
    \begin{subfigure}{\linewidth}
        \includegraphics[width = \linewidth]{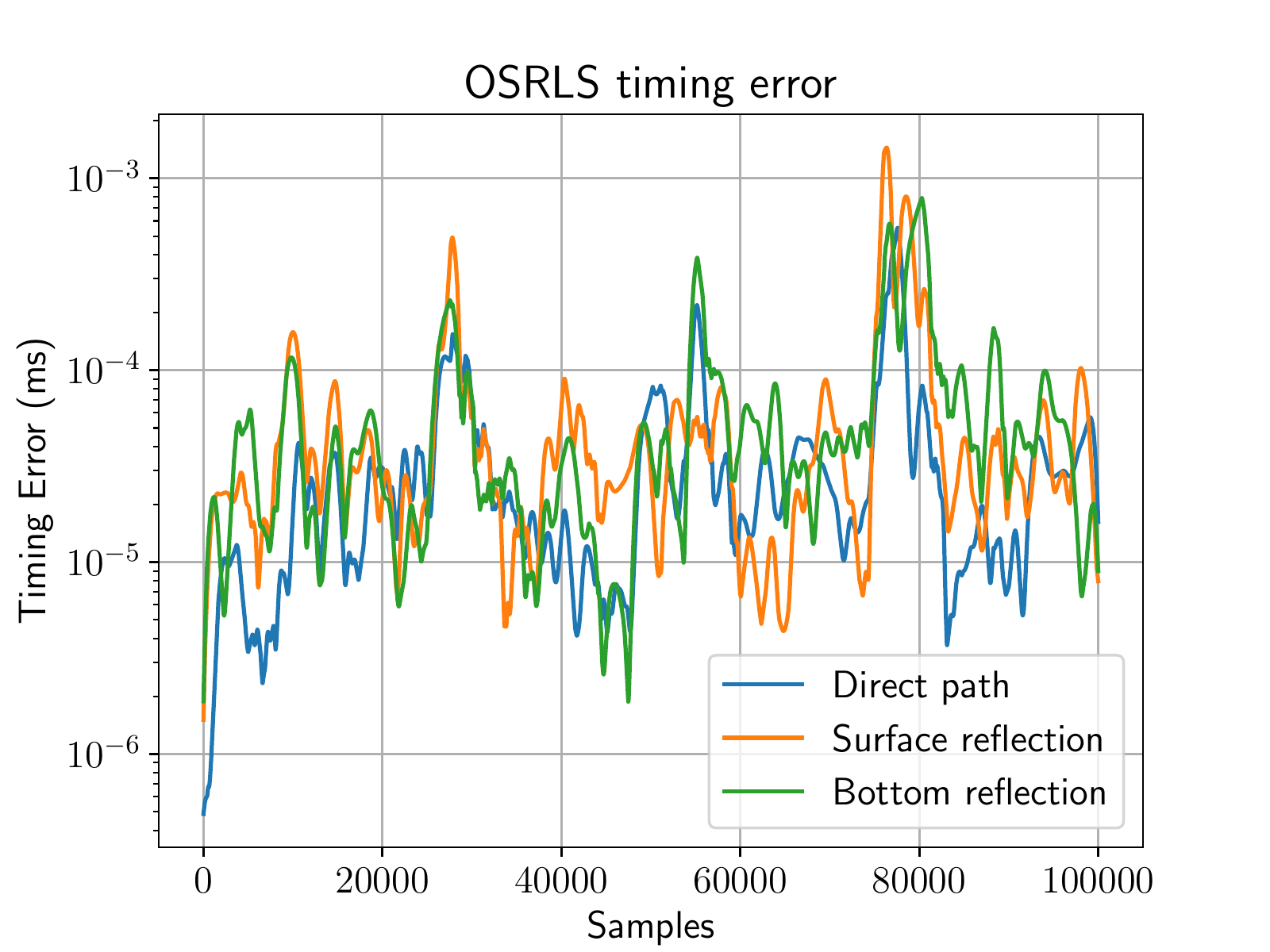}
    \end{subfigure}
    \begin{subfigure}{\linewidth}
        \includegraphics[width = \linewidth]{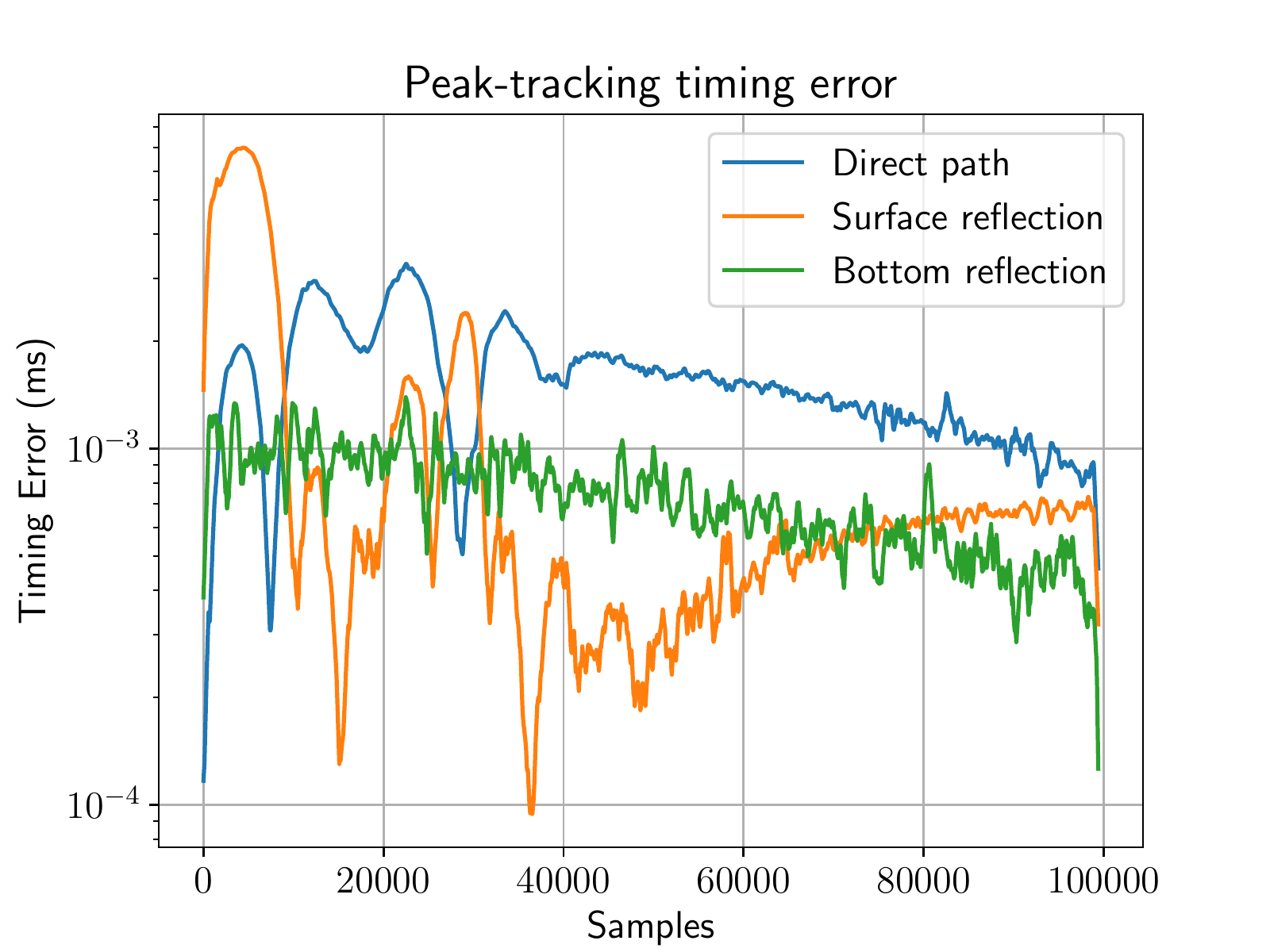}
    \end{subfigure}
    \caption{Sample timing error recovered from the estimated Doppler factor using OSRLS (top) and peak-tracking (bottom) algorithms with sample rate of 200~kHz. The absolute valued errors are averaged over 1000 samples. }\label{fig:err}
\end{figure}
Timing estimates of samples are recovered using estimated Doppler for OSRLS and peak delay for the peak-tracking method. The errors between the ground truth and estimated values are shown in Fig. \ref{fig:err} for both OSRLS and peak-tracking methods. Given that the sample rate is 200~kHz, the timing error of 0.0025~ms could result in a missed sample. It can be observed that the timing estimation error of the OSRLS-based approach is always within a sample interval for this simulation, while the peak detection shows occasional sample misses for the direct and bottom reflections, and a bias larger than a sample interval for the surface reflection after approximately 25000 iterations.

\section{Conclusion} \label{sec:con}
We presented an OSRLS-based sequential method, inspired by SLS, for time-varying non-uniform Doppler estimation. We formulated the problem in an SLS framework by using a first-order Taylor expansion with respect to the change in the Doppler factor. Further, we enhanced the numerical stability of the algorithm through the use of a gradient perturbation. The proposed algorithm was tested in simulation, and has shown substantially improved performance as compared to the peak-tracking method. Future work will be focused on incorporating the OSRLS-based Doppler estimation algorithm into the underwater communication framework. The first step will be evaluating the algorithm with experimental measurements and incorporation of joint Doppler and channel gain estimation. In addition to the performance of the proposed method with known transmission signal which is demonstrated in this paper, further study will be made on the performance under the decision-driven mode of an acoustic communication system.

\bibliographystyle{IEEEtran}
\IEEEtriggeratref{7}
\bibliography{IEEEabrv,reference}

\end{document}